\documentclass[pre,preprint,aps,eqsecnum,raggedbottom]{revtex4}
\usepackage{graphicx}
\usepackage{epsf}
\usepackage{epstopdf}

\newcommand{\e}{{\rm e}}
\newcommand{\ep}{\epsilon}

\newcommand{\bet}{\beta}

\newcommand{\bea}{\begin{eqnarray}}
\newcommand{\eea}{\end{eqnarray}}
\newcommand{\be}{\begin{equation}}
\newcommand{\ee}{\end{equation}}
\newcommand{\ba}{\begin{eqnarray}}
\newcommand{\ea}{\end{eqnarray}}

\newcommand{\nn}{\nonumber}
\newcommand{\la}{\label}

\newcommand{\pa}{\partial}

\def\t1{e_{_T}}
\def\v1{e_{_V}}

\begin{document}
%\tightenlines
\title{Parafermions in plain sight}

\author{Siu A. Chin}
\email{chin@physics.tamu.edu.}
\affiliation{Department of Physics and Astronomy,
Texas A\&M University, College Station, TX 77843, USA}

\author{A. Chaudhary}
\email{aarifchaudharyg@gmail.com}
\affiliation{Hendrix Industries, Sealy, Texas 77474, USA}

%\date{\today}
\begin{abstract}

We show that for any potential with a discrete energy spectrum, the well-known interpolation between ideal boson and fermion partition functions at discrete values of $\xi=-1/m$ yielded zero temperature ground state energies corresponding to $m$ fermions occupying a single quantum state. The grand canonical partition function in this case can be a result from genuine parastatistics.

\end{abstract}
\maketitle

\section {Introductions}

In 1953, Green\cite{gre53} showed that 
``spin-half fields can be quantized in such a way that an arbitrary finite number of particles can exist in each eigenstate". 
He refers to this as ``the generalization of the statistics".
A decade later, Greenberg\cite{gre64} suggested that the spin-3/2 $\Delta^{++}$ baryon can be composed of three spin-1/2 u quarks
in the same s-state, if the latter were parafermions of order $p=3$, 
allowed by Green's ``parastatistics" for three fermions to occupy the same quantum state. 
However, Greenberg clearly stated that the parafermion's annihilation and creation operators, $a^{(\alpha)}$ and $a^{(\beta)+}$,
``satisfy the anticommutation rules for the same $\alpha$, and commute for different $\alpha$ and $\beta$".
 These are the same commutation rules obeyed by conventional fermions having internal states labeled by $\alpha$ and $\beta$.  Greenberg and Messiah\cite{gre65} confirmed this latter interpretation and history has followed it in adopting the view that quarks have three internal ``color" states, coupled to $SU(3)$ gauge fields\cite{han65,fri72,fri73}. 
 
Due to the ``conventionality argument"\cite{bak15}, Green's theory of parafermions has been shown to be equivalent to conventional fermions with internal states, preventing experimental observation of
genuine parastatistics.
However, recently Wang and Hazzard\cite{wan25,wan26} have suggested that there might be quasiparticle excitations with $R$-parastatistics that cannot be reduced to conventional statistics of bosons or fermions. 
They also pointed out that different exchange statistics may produce the same partition function, 
such as their examples 1, 2 and 6.

In this work, we show that for $N$ quantum particles in any potential with a discrete spectrum,
the well-known $\xi$-interpolation between ideal boson and fermion partition functions
at $\xi=-1/m$, yielded zero temperature ground state energies corresponding to the occupations 
of $m (<N)$ fermions in a single energy state. This is a complete surprise, since the recursion relation 
(\ref{zn}) below has been known for a long time, but no one had noticed this until this work.
As shown below, the $m$-particle occupation is consistent with parafermions of the simplest kind,
that of ordinary fermions with $m$-internal states. However, since the canonical partition function 
cannot always distinguishes different exchange statistics, this $\xi=-1/m$ partition function can also 
corresponds to non-trivial exchange of $R$-paraparticles. We will return to this point in the Conclusion.

\section {The intuitive picture}
\la{ip}

It has been known for sometime\cite{for71,bor93,bro97} that the partition function $Z_N$ for $N$ non-interacting 
bosons and fermions is given by the recursion relation (Newton's identity),
\be
Z_N=\frac1{N}\sum_{k=1}^{N}\xi^{k-1}z_k Z_{N-k},
\la{zn}
\ee 
where $\xi=1$ for bosons, $\xi=-1$ for fermions, and $Z_0=1$.
Recently, it has been suggested\cite{xio22} that by extrapolating from $\xi=1$ to $\xi=-1$,
one might be able the alleviate the sign problem\cite{comment} in fermion path integral Monte Carlo.
This work is not about extrapolating $\xi$, but about the behavior of the partition function $Z_N$ at all values of $\xi$.

If $\xi$ is viewed as the weight factor multiplying the wave function under the exchange of
any two indistinguishable particles, then upon the second exchange, the wave function is restored to its original form
and one must have $\xi^2=1$. Therefore any {\it real} $\xi\ne\pm 1$ is non-unitary. Particles with such
a exchange weight are obviously unphysical and therefore rightly ``fictitious"\cite{xio22}.

However, in this work, we will offer a different perspective on $\xi$. We will regard (\ref{zn}) simply
as a mathematical model for the partition function $Z_N$ with a single parameter $\xi$. All real values 
of $-\infty<\xi<\infty$ are mathematically legitimate.  
At each value $\xi$, we will simply let $Z_N$ tell us, what the $N$ particles are doing and whether or not
they are physical.

In this, and the next section, we will show that, in the zero temperature limit, 
{\it all} finite values of $\xi$
converge to the energy of $N$ bosons, except for a set of discrete values $\xi=-1/m$ with $m=\{1,2,3\cdots (N-1)\}$.
This is true for any potential with a discrete spectrum, but here, 
we will numerically illustrate this using the harmonic oscillator potential.

For a $d$-dimension harmonic oscillator, the energy spectrum is given by
\ba
E&=&(n_1+\frac12)+(n_2+\frac12)+\cdots (n_d+\frac12),
% &=&\sum_{i=1}^d n_i+\frac{d}{2},
\ea
where $n_i=\{0,1,2,\cdots \infty\}$. The one-particle partition function $z_1$ is given by
\ba
z_1(\tau)&=&\sum_{n_1}\sum_{n_2}\cdots\sum_{n_d}\e^{-\beta E},\la{dsum}\\
&=&\left( \frac{b^{1/2}}{1-b}\right)^d,
\ea
where $b=\e^{-\beta}$ and $\beta=1/T$ is the inverse temperature. The $z_k$ partition function in (\ref{zn}) is defined by 
\ba
z_k&=&z_1(k\tau)=\left( \frac{b^{k/2}}{1-b^k}\right)^d.
\ea
For a particularly simple spectrum, we will consider only the case of $d=2$ in this section.
The general potential case, not necessary harmonic, will be considered in Sect.\ref{fd}.

Given $Z_N$ from (\ref{zn}), the energy can be computed from
\be
E=-\frac{(\pa Z_N/\pa \bet)}{Z_N}=\frac{b(\pa Z_N/\pa b)}{Z_N}
\ee

For $N=2$ particles, the partition function is
\ba
Z_2&=&\frac1{2!} (z_1^2+\xi z_2),\nn\\
&=&\frac1{2!} \left(\frac{b^2}{(1-b)^4}+\xi\frac{b^2}{(1-b^2)^2} \right),
\la{z2}
\ea
and the resulting energy for various values of $\xi$ is plotted as a function of $\beta$  in Fig.{\ref{n2xt}}.

One sees that at the zero-temperature limit of $\bet\rightarrow\infty$, $b\rightarrow 0$,
$\xi=-1$ correctly converges to the fermion energy limit of $E=3$ whereas
{\it all} $\xi\ne -1$ energies collapse to the boson limit of $E=2$.

To understand these behaviors, one can expand $Z_2$  (ignoring the constant $1/2!$)
in the limit of $b\rightarrow 0$, giving,
\ba
Z_2&=&(1+\xi)b^2+4b^3+(10+2\xi)b^4 +\cdots\nn\\
b\frac{\pa (Z_2)}{\pa b}&=&2(1+\xi)b^2+4(3 b^3)+4(10+2\xi)b^4+\cdots.
\ea
The derivative simply multiplies each term by the exponent of $b$, giving the energy,
\be
E_2(\tau,\xi)=\frac{2(1+\xi)+3(4b)+4(10+2\xi)b^2+\cdots}
{(1+\xi)+4b+(10+2\xi)b^2 +\cdots}.
\ee
Therefore, as long as $\xi\ne -1$, in the limit of $b\rightarrow 0$, the energy is always 2,
regardless the actual value of $\xi$.
Only in the case of $\xi=-1$, when the first term vanishes, that the energy is 3. 

\begin{figure}[t]
	\includegraphics[width=0.70\linewidth]{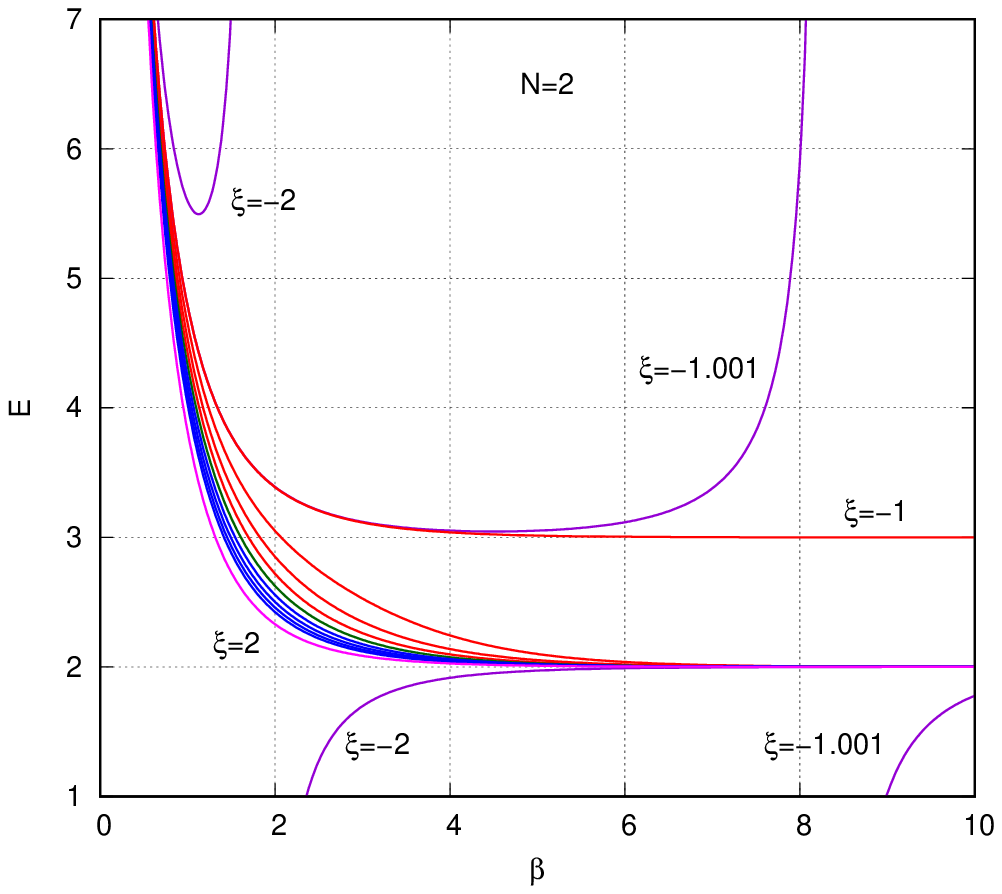}
	\caption{ (color online) 
		The energy of two non-interacting particles in a two dimensional harmonic potential,
		from top to bottom, at $\xi=-2$, $-1.001$, $-1$, $-3/4$, $-1/2$, $-1/4$,
		0, 1/4, 1/2, 3/4, 1 and 2. Red and blue lines corresponds to negative and positive $\xi$ values.
		Purple is used for $\xi$ outside of the range [-1,1] and green for $\xi=0$.}
	\la{n2xt}
\end{figure}

However,  when $\xi<-1$, the partition function can vanish, from (\ref{z2}), at
\be
\e^{-\bet}=\frac{\sqrt{-\xi}-1}{\sqrt{-\xi}+1},
\ee 
and shows up as a pole in the energy, as shown for $\xi=-2$ and $\xi=-1.001$ in Fig.\ref{n2xt}.
Since a negative partition function is unphysical, one can rightly dismiss the mathematical model at $\xi<-1$ 
as unphysical. Moreover, one can easily tell that these $\xi$ values are unphysical 
because their energies converge to the boson energy from {\it below}, which is physically impossible.

\begin{figure}[t]
	\includegraphics[width=0.70\linewidth]{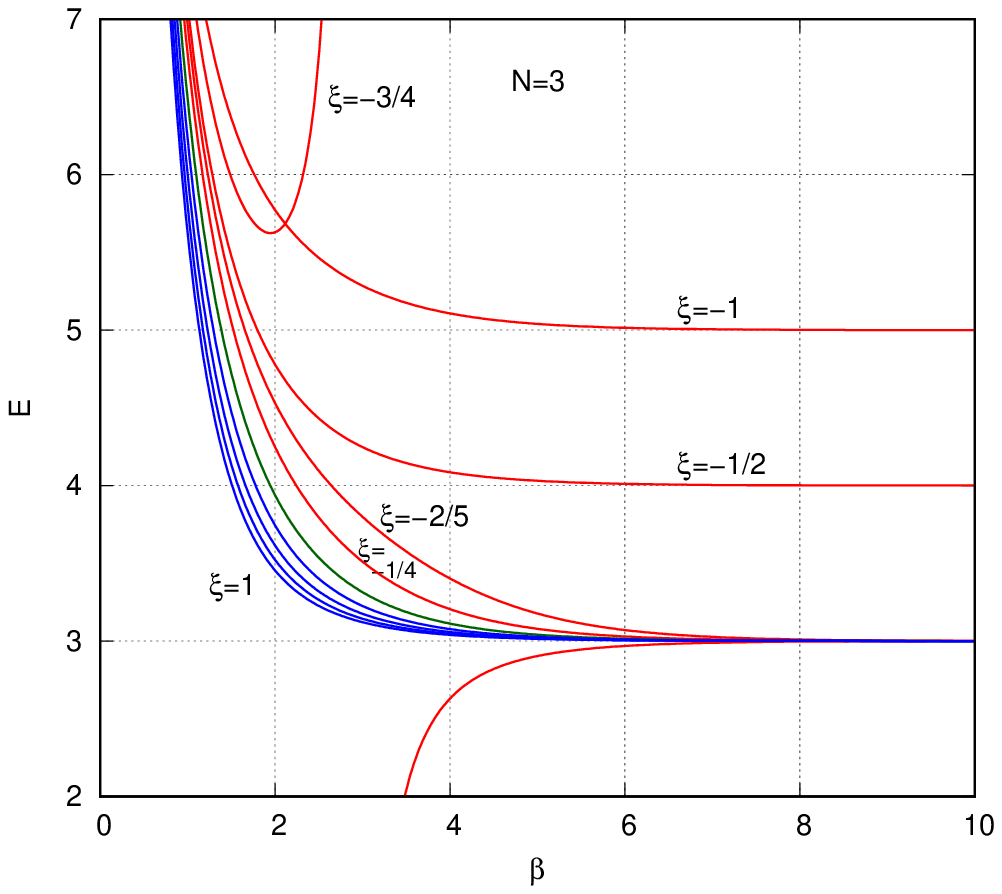}
	\caption{ (color online) 
		Similar to Fig.\ref{n2xt} for $N=3$ particles.}
	\la{n3x}
\end{figure}

For $N=3$ particles, one has
\be
Z_3=\frac1{3!}(z_1^3+3\xi z_2z_1+2\xi^2z_3),
\la{z3}
\ee
and similar expansions give the energy
\be
E_3(\tau,\xi)=\frac{3(1+\xi)(1+2\xi)+24(1+\xi)b+15(7+5\xi)b^2+\cdots}
{(1+\xi)(1+2\xi)+6(1+\xi)b+3(7+5\xi)b^2 +\cdots},
\la{e3}
\ee
which dictates that, unless
$\xi=-1$ which converges to $E_3=5$, or $\xi=-1/2$ which gives $E_3=4$, all other values of $\xi$
collapse to the boson energy of $E_3=3$ in the zero temperature limit, as shown Fig.\ref{n3x}.
Since (\ref{z3}) is quadratic in $\xi$, there are also more unphysical poles, one of which is at $\xi=-3/4$,
also shown in Fig.\ref{n3x}.

\begin{figure}[t]
	\includegraphics[width=0.70\linewidth]{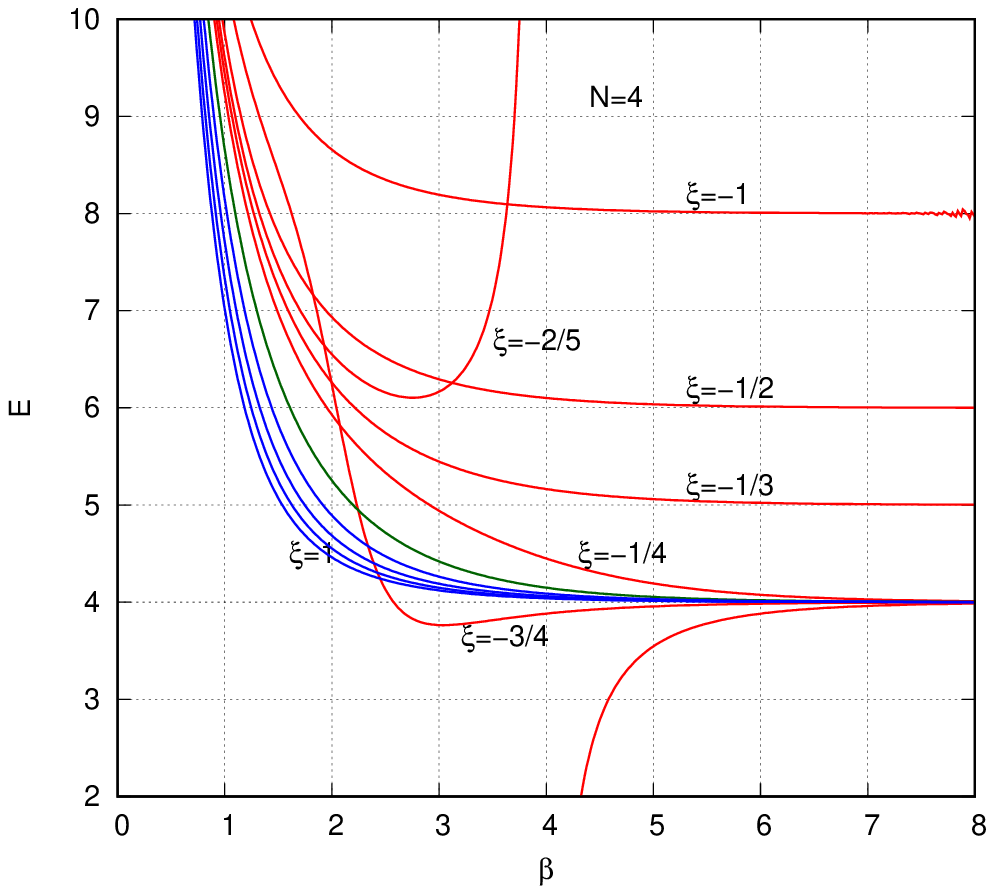}	
	\caption{ (color online) 
			Similar to Fig.\ref{n2xt} for $N=4$ particles.}
	\la{n4x}
\end{figure}

\begin{table}
	\caption{Energies for $N$ particles in a 2D harmonic oscillator at various values of $\xi$.}
	\begin{center}
		\begin{tabular}{c r r r r }
			\colrule
			$N$ & \qquad $\xi$  &\qquad E \qquad & sum\qquad  &\qquad occupation\\
			\colrule
			2 &\qquad -1    &\qquad 3 \qquad &1+2\qquad    & 1,1,0,...\qquad   \\
			&1     &\qquad 2 \qquad  &\qquad 1+1\qquad   & 2,0,0,...\qquad  \\
			\colrule
			3 &-1    &\qquad 5 \qquad   &\qquad 1+2+2  &1,1,1,0,...\quad   \\
			&-$\frac12$  &\qquad 4 \qquad  &\qquad 1+1+2  &2,1,0,0,...\quad   \\
			&1     &\qquad 3 \qquad  &\qquad 1+1+1  &3,0,0,0,...\quad   \\
			\colrule
			4 &-1    &\qquad 8 \qquad    &\qquad 1+2+2+3  &1,1,1,1,0,...\quad  \\
			&-$\frac12$  &\qquad 6 \qquad    &\qquad 1+1+2+2  &2,2,0,0,0,...\quad   \\
			&-$\frac13$  &\qquad 5 \qquad    &\qquad 1+1+1+2  &3,1,0,0,0,...\quad  \\
			&1     &\qquad 4 \qquad    &\qquad 1+1+1+1  &4,0,0,0,0,...\quad   \\
			\colrule
		\end{tabular}
	\end{center}
	\label{tab1}
\end{table}

For $N=4$ particles, one has
\be
Z_4=\frac1{4!}(z_1^4+6\xi z_2z_1^2+3\xi^2 z_2^2 +8\xi^2z_3z_1+6\xi^3 z_4).
\la{z4}
\ee
The resulting energy
\be
E_4(\tau,\xi)=\frac{4c_0+5c_1b+6c_2 b^2 +7c_3 b^3+8c_4 b^4+\cdots}
{c_0+c_1b+c_2 b^2 +c_3 b^3+c_4 b^4+\cdots},
\ee
where
\be
c_0=(1+\xi)(1+2\xi)(1+3\xi),\quad c_1=8(1+\xi)(1+2\xi),\quad c_2=36(1+\xi)^2,
\ee
\be
c_3=24(1+\xi)(5+2\xi),\quad c_4=(330+348\xi+102\xi^2+12\xi^3),
\ee
now dictates that
$\xi=-1$ converges to $E_4=8$, $\xi=-1/2$ to $E_4=6$, $\xi=-1/3$ to $E_4=5$
and all other values of $\xi$ collapse to the boson energy of $E_4=4$ (with unphysical convergence from below
at $\xi=-2/5$ and -3/4).

These examples demonstrate that, for $N$ particles in the limit of low temperature $b\rightarrow 0$, the partition function must have
the form
\be
Z_N=c_0b^N+c_1 b^{N+1}+c_2b^{N+2}+\cdots,
\ee
so that
\be
E_N(\tau,\xi)=\frac{Nc_0+(N+1)c_1 b+(N+2)c_2 b^2+\cdots}
{c_0+c_1b+c_2 b^2+\cdots}.
\ee
For all $\xi$ such that $c_0\ne 0$, the energy will collapse to the boson limit of $E_N=N$ in the zero temperature limit.
Non-boson energy is possible only if $c_0=0$.
The above examples suggest that (restoring the $1/N!$ constant)
\be
c_0=\frac1{N!}(1+\xi)(1+2\xi)(1+3\xi)...(1+(N-1)\xi)
\la{czero}
\ee
and these higher energies occurs at discrete
values of $\xi$:
\be
\xi=\left\{-1,-\frac12,-\frac13,\cdots -\frac1{(N-1)}\right\}.
\la{xiv}
\ee
An formal derivation of (\ref{czero}) will be given in the next section.

To appreciate the significance of these higher energies at zero temperature, 
we list them for each value of $\xi$ in Table \ref{tab1}.
For each $N$ values, the $\xi=1$ energy is the sum of $N$ ground state energy of $E_0=1$, corresponding
to the occupation of $N$ bosons in the ground state. The $\xi=-1$ energies correspond to a single occupation of
each successive energy level, befitting of fermions. For $\xi=-1/2$, at $N=3$ the energy is a 
double occupation of the ground state and one particle in the excited state. 
At $N=4$, it is the double occupation of both the ground state
and the first excited state. 
For $\xi=-1/3$ at $N=4$, it is the triple occupation of
the ground state plus one excited state. 
These examples
show that, for $2 \le m \le (N-1)$, $N$ particles at $\xi=-1/m$ behave like Green or Greenberg's parafermions of order $m$,
consistent with ordinary fermions having $m$ internal states. 
We will formally derived this for a general potential in the next Section.

\section {The formal derivation}
\la{fd}

The derivation of the grand canonical partition function corresponding to (\ref{zn})
\be
\Xi(t)=\sum_{N=0}^{\infty} Z_N t^N,
\ee
is a standard exercise in algebraic combinatorics\cite{mac95,sta24}. We outline the essential steps here for completeness.
Summing over both sides of (\ref{zn}) gives
\ba
\sum_{N=1}^{\infty} N Z_Nt^N&=&\sum_{N=1}^{\infty}\sum_{k=1}^{N}\xi^{k-1}z_k Z_{N-k}t^N,\nn\\
t\frac{d\Xi}{dt}&=&\sum_{N=1}^{\infty}\sum_{k=1}^{N}\xi^{k-1}z_k Z_{N-k}t^N.
\ea
By defining $j=N-k$, with $j\ge 0$ and therefore $N=j+k$, the double sums on the RHS can be disentangled into two independent sums:
\ba
t\frac{d\Xi(t)}{dt}&=&\left(\sum_{k=1}^{\infty}\xi^{k-1}z_kt^k\right)\left(\sum_{j=0}^{\infty} Z_jt^j\right),\nn\\
&=&\left(\sum_{k=1}^{\infty}\xi^{k-1}z_kt^k\right)\Xi(t),\nn\\
\frac{d\ln\Xi(t)}{dt}&=&\sum_{k=1}^{\infty}\xi^{k-1}z_kt^{k-1}.
\ea
Integrating both sides gives
\ba
\ln\Xi(t)&=&\sum_{k=1}^{\infty}\frac{\xi^{k-1}}{k}z_kt^k,
\ea
and therefore
\be
\Xi(t)=\exp\left(\sum_{k=1}^{\infty}\frac{\xi^{k-1}}{k}z_kt^k\right).
\la{gcx}
\ee

Consider now a general potential in any dimension with a discrete energy spectrum $\ep_n$ and degeneracy $g_n$
labeled by $n=0,1,2\cdots\infty$, then
\ba
z_1(\tau)&=&\sum_{n}g_n\e^{-\beta \ep_n},\nn\\
z_k(\tau)&=&z_1(k\tau)=\sum_ng_n\e^{-k\beta \ep_n}.
\ea
Substitute the above $z_k$ into (\ref{gcx}) gives
\ba
\Xi(t)&=&\exp\left(\sum_{k=1}^{\infty}\frac{\xi^{k-1}}{k}\sum_ng_n\e^{-k\beta \ep_n} t^k\right)
=\exp\left(\sum_n \frac{g_n}{\xi}\sum_{k=1}^{\infty}\frac{1}{k}(\xi\e^{-\beta \ep_n} t)^k\right),\nn\\
&=&\exp\left(-\sum_n \frac{g_n}{\xi}\ln(1-\xi\e^{-\beta \ep_n} t) \right)
=\prod_n \left(1-\xi b^{\ep_n} t\right)^{-g_n/\xi}.
\la{gn}
\ea

In the low temperature limit of $b\rightarrow 0$, it is only necessary to keep the $n=0$ term,
\ba
\Xi(t)&=&\left(1-\xi x \right)^{-g_0/\xi}.
\ea
where we have defined $x=b^{\ep_0}t$.
Compare this to the generalized binomial expansion
\be
(1-x)^{-a}=\sum_{k=0}^{\infty}\frac{R_k(a)}{k!}x^k
\ee
where
$R_k(a)=a(a+1)(a+2)\cdots(a+(k-1))$, one has
\ba
\Xi(t)
&=&\sum_{k=0}^{\infty}\frac{R_k(g_0/\xi)}{k!}(\xi x)^k
\ea
The coefficient of the $t^N$ term is therefore $b^{N\ep_0} c_0$, with $N$-boson energy $N\ep_0$ and
\ba
c_0&=&\frac{R_N(g_0/\xi)}{N!}\xi^N,\nn\\
&=&\frac1{N!}\frac{g_0}{\xi}\left(\frac{g_0}{\xi}+1\right)\left(\frac{g_0}{\xi}+2\right)\cdots\left(\frac{g_0}{\xi}+(N-1)\right)\xi^N,\nn\\
&=&\frac{g_0}{N!}\prod_{j=1}^{N-1}(g_0+j\xi),
\ea
which is (\ref{czero}) for $g_0=1$. Since $c_0$ follows only from keeping the $n=0$ term, 
it is a {\it universal} coefficient, same for any potential with a discrete spectrum having the same ground state
degeneracy $g_0$. For our constant references to the harmonic oscillator, we will assume $g_0=1$ unless stated otherwise.

When $\xi=-1/m$, the grand partition function is
\ba
\Xi(t)&=&\left[\prod_n \left(1+\e^{-\beta \ep_n} \frac{t}{m}\right)^{g_n}\right]^m,
\la{xim}
\ea
which is $m$ copies of the $\xi=-1$ grand canonical partition function allowing
$m$ parafermions to occupation of each state. The resulting canonical partition function, the coefficient of $t^N$, is therefore
\be
Z_N^{(-1/m)}=m^{-N}Z_{N}^{(m)}
\ee
where $Z_{N}^{(m)}$ is the $N$-fermion partition function allowing $m$-state occupation.

To see how this $m$-state occupation actually works out, consider the one dimensional case without
the complication of degeneracy with $g_j=1$ and $\ep_j=j+1/2$:
\ba
\Xi(t)&=&\left[\prod_{j=0}^\infty \left(1+b^j x\right)\right]^m,
\ea
with $x=b^{1/2}t/m$.
From the q-binomial identities, the coefficient of the $x^n$ term for $m=1$ is 
\ba
[x^n]\prod_{j=0}^\infty \left(1+ b^jx \right)&=&\frac{b^{n(n-1)/2}}{(1-b)(1-b^2)\cdots (1-b^n)},\nn\\
\lim_{b\rightarrow 0}&\rightarrow& b^{n(n-1)/2}
\ea
and therefore the coefficient of $t^n$ is
\be
Z_n(b)\rightarrow b^{n(n-1)/2}b^{n/2}=b^{n^2/2}.
\ee
For $m\ne 1$, in the limit of $b\rightarrow 0$, (ignoring the $1/m$ factor multiplying $t$) one then has
\ba
\Xi(t)&=&\left[\sum_{k=0}^\infty b^{k^2/2}t^k\right]^m,\nn\\
&=& \sum_{k_1=0}^\infty\sum_{k_2=0}^\infty\cdots \sum_{k_m=0}^\infty b^{(\sum_{i=1}^m k_i^2/2)}\,t^{(\sum_{i=1}^m k_i)}.
\ea
For the simplest illustration, consider the case where $N$ is a multiple of $m$, then the minimum power of $b$ subject to the constraint 
$\sum_{i=1}^m k_i=N$ is for each $k_i=N/m$ and therefore
\be
Z_N^{(m)}\rightarrow b^{m(N/m)^2/2},
\ee
with ground state energy
\be
E_N= m\, \frac12\left(\frac{N}{m}\right)^2,
\ee
which is $m$ times the energy of occupation to level $(N/m)$.

\section {Conclusions}

In this work, we have shown, both by numerical examples and by formal derivations, that for
$\xi=-1/m$, with $2\le m \le(N-1)$, the $N$ particles in the canonical partition function (\ref{zn}) 
are consistent with being fermions with $m$-internal.
However, Wang and Hazzard\cite{wan26} have pointed out that for a single mode or energy state, 
the canonical partition function is the same for their example 1, 2, and 6 statistics, where example 1 corresponds to ordinary fermions having $m$ internal states. Since the particles are noninteracting, in the case of multiple modes, the grand canonical partition function (GCPF) would then be a product over all states as in (\ref{xim}). In this case, examples 2 and 6, with non-trivial exchange parastatistics, would produce the same GCPF as our (\ref{xim}) with $m$ internal states. 
However, since non-trivial exchange does not automatically imply genuine parastatistics, 
equivalence with examples 2 and 6 does not necessarily mean that (\ref{xim}) is a product of parastatistics.

To make this final connection ironclad, 
we note that any $R$-matrix satisfying the conditions of their classification theorem IV.1\cite{wan26} with $\theta = -1$, must all have the same GCPF (\ref{xim}), include those with genuine parastatitics. In particular, for $N\ge4$ and $\xi=-1/4$, the GCPF of (\ref{xim}) corresponds to their example 5,
which Wang and Hazzard\cite{wan26} have characterized as ``physically (observably) 
distinct from fermions and bosons".
Therefore, (\ref{xim}) at $\xi=-1/4$ produces a GCPF from genuine parastatistics.

%%%%%%%%%%%%%%%%%%%%%%%%%%%%%%%%%%%%%%%%%%%%%%%%%%%%%%%

\begin{thebibliography}{99}
\bibitem{comment}
Parameter extrapolation as a mean of overcoming the sign problem is an old idea tried 30+ years ago\cite{alh94},
but was eventually abandoned with the realization that extrapolation over a parameter gap is mathematically impossible without first knowing the answer.
\end{thebibliography}
\end{document}